\documentclass[reprint, prb]{revtex4-2}
\usepackage{graphicx} 
\usepackage{amsmath}
\usepackage{amsfonts}
\usepackage{chemformula}
\usepackage{siunitx}
\usepackage{bm}
\usepackage{multirow}
\usepackage{lipsum}

\DeclareSIUnit{\angstrom}{\textup{\AA}}

\begin{document}
\title{Choosing Tight-Binding Models for Accurate Optoelectronic Responses}

\author{Andreas Ghosh}
\affiliation{Department of Chemistry, University of Pennsylvania, Philadelphia, Pennsylvania 19104--6323, USA}
\author{Aaron M. Schankler}
\affiliation{Department of Chemistry, University of Pennsylvania, Philadelphia, Pennsylvania 19104--6323, USA}
\author{Andrew M. Rappe}
\affiliation{Department of Chemistry, University of Pennsylvania, Philadelphia, Pennsylvania 19104--6323, USA}

\date{\today}

\begin{abstract}

Tight-binding models provide great insight and are a low-cost alternative to \emph{ab initio} methods for calculation of a material's electronic structure. These models are used to calculate optical responses, including nonlinear optical effects such as the shift current bulk photovoltaic effect. The validity of tight-binding models is often evaluated by comparing their band structures to those calculated with Density Functional Theory.
However, we find that band structure agreement is a necessary but not sufficient condition for accurate optical response calculations.
In this Letter, we compute the shift current response and dielectric tensor using a variety of tight-binding models of \ch{MoS2}, including both Slater-Koster and Wannier tight-binding models that treat the Mo $4d$ orbitals and/or S $3p$ orbitals.
We also truncate hoppings in the Wannier function models to next-nearest neighbor, as is common in tight-binding methods, in order to gauge the effect on optical response.
By examining discrepancies in energies and optical matrix elements, we determine the interpolation quality of the different tight-binding models and establish that agreement in both band structure and wavefunctions is required to accurately model optical response, 
\end{abstract}

\maketitle

\section{Introduction}

Tight-binding methods have a long and successful history in condensed matter physics.
At their best, these models can capture the salient electronic physics of a system at significantly reduced cost, enabling conceptual understanding of responses as well as calculations of larger systems and more complex interactions than would be possible using fully \emph{ab initio} methods. 
Due to their relative simplicity, tight-binding models provide more straightforward routes to extracting physical insight from numerical results~\cite{Goringe97p1447}.
These models have enabled great strides in the study of electron transport, low-dimensional materials, and topological properties.

Tight-binding methods have proven invaluable for studying nonlinear optical properties, both in calculating optical spectra of materials and in developing structure-property understanding~\cite{Tan16p16026,Ishizuka17p033015}.
Here, we highlight calculation of the shift photocurrent, one of the main mechanisms of the bulk photovoltaic effect (BPVE). The BPVE refers to a DC photocurrent that is generated by uniform illumination of homogeneous materials lacking inversion symmetry~\cite{Belinicher80p199,vonBaltz81p5590}. In the shift current, the amplitudes of conduction band wavefunctions interfere asymmetrically, resulting in a net real space displacement of electrons on photoexcitation~\cite{Tan16p16026}.
Though there are several effective schemes for calculating the shift current response using \emph{ab initio} calculations~\cite{Sipe00p5337,Young12p116601,Ibanez18p245143}, tight-binding treatments have also yielded important insights about the shift current mechanism.
These include links to other material properties~\cite{Tan16p16026,Fregoso17p075421,Tan19p085102}, a detailed picture of the time- and power-dependence of the response~\cite{Rajpurohit22p094307,Bajpai19p025004}, and new understanding of the shift current response to magnetic field~\cite{Dai23pl201201}.
Given the great explanatory power of tight-binding models, methods to parameterize tight-binding models that can accurately provide the optical properties of real materials is of central importance.

In this Letter, we compare the accuracy of the shift current response for monolayer \ch{MoS2} calculated using different types of tight-binding models.
We show that accurately reproducing the Density Functional Theory (DFT) band structure is often not enough to ensure that optical responses such as the shift current and the dielectric tensor are also reproduced accurately.
We argue that Wannier tight-binding models capture the optoelectronic response of \ch{MoS2} more accurately than Slater-Koster symmetry-based tight-binding models.
Many tight-binding models have been reported for transition-metal dichalcogenides (TDMs), such as the three-band model presented in Ref.~\onlinecite{Liu13p085433} and the extended five-band version presented in Ref.~\onlinecite{Wu15p075310}. Both of these models are parameterized using the Slater-Koster method~\cite{Slater54p1498}, and they faithfully represent the main features of the DFT band structure; however, these models do not capture optical properties as accurately.
Wannier interpolation has been used to facilitate dense reciprocal space sampling in \emph{ab initio} calculations~\cite{Ibanez18p245143}. Because of their localization, Wannier functions can also be readily reduced to a tight-binding description.
We find that such models are an attractive middle ground that can capture hybridization and delocalization in the wavefunctions of interest, which are crucial for calculation of optical properties, while still maintaining the advantageous scaling behavior and good interpretability of tight-binding models.

To demonstrate this, we construct two Wannier function models of \ch{MoS2}---one that contains only Mo $4d$ orbitals (five-band) and another with both Mo $4d$ and S $3p$ orbitals (eleven-band).
We compare the band structures and both the linear and nonlinear optical responses of these models to those of the Slater-Koster tight-binding models proposed in Refs.~\cite{Wu15p075310,Liu13p085433} and to fully \emph{ab initio} treatments.
We also consider simplifications to the Wannier tight-binding model, including restricting the range of the hopping terms, allowing us to assess the relatively short hopping range of typical tight-binding models.
These comparisons show that accurately reproducing the DFT band structures is not a guarantee that optical properties will match the first-principles results.
We argue that because optical properties depend directly on the wavefunction character across the Brillouin zone, tight-binding models fit to reproduce wavefunctions, in addition to band structures, are necessary for the most accurate results.
This insight will guide the development of improved model systems that can facilitate calculations of ultrafast and magneto-optical responses.

\section{Methods}

To provide an \emph{ab initio} reference, we perform density functional theory (DFT) calculations of the 2H phase of monolayer \ch{MoS2} using Quantum Espresso~\cite{Giannozzi09p395502,Giannozzi17p465901}. To minimize inter-layer interaction, \qty{16}{\angstrom} of vacuum was included between periodic images. The calculation employed the PBE exchange-correlation functional~\cite{Perdew96p3865} and optimized norm-conserving pseudopotentials generated using the OPIUM package~\cite{Rappe90p1227,Ramer99p12471}.
The charge density was computed self-consistently on a \numproduct{12x12x1} $k$-point grid, and the wavefunctions were calculated on a denser \numproduct{24x24x1} mesh to compute the shift current and other optical properties.
The shift current susceptibility $\sigma_{rsq}$ was calculated according to the procedure outlined in Ref.~\onlinecite{Young12p116601},
\begin{multline}
\label{eq:shift}
\sigma_{rsq}(\omega) = \frac{2\pi e^3}{\hbar \omega^2}
    \sum_{nm} \int_{BZ} \frac{d \bm{k}}{(2 \pi)^3}
    \left(f_{m\bm{k}} - f_{n\bm{k}}\right) \\
    \times v_{nm}^s(\bm{k}) v_{mn}^r(\bm{k})
    \left[ -\frac{\partial \phi_{nm}^r(\bm{k})}{\partial k_q} 
    - \left[A^q_{mm}(\bm{k}) - A^q_{nn}(\bm{k})\right]
    \right]\\
    \times
    \delta\left(E_{n\bm{k}} - E_{m\bm{k}} - \hbar \omega\right) 
\end{multline}
where $nm$ sum over bands, and $rsq$ index Cartesian coordinates. Here $f_{n\bm{k}}$, $E_{n\bm{k}}$, and $v_{mn}^r(\bm{k})$ denote band fillings, energies, and velocities respectively, $\phi_{mn}^r$ is the phase of the corresponding momentum matrix element, and $A^q_{nn}$ is the Berry connection.
In this Letter, we compute the $yyY$ element of the shift current tensor and the imaginary part of the $xx$ dielectric tensor to show calculations in both Cartesian directions.

\begin{figure}
    \centering
    \includegraphics{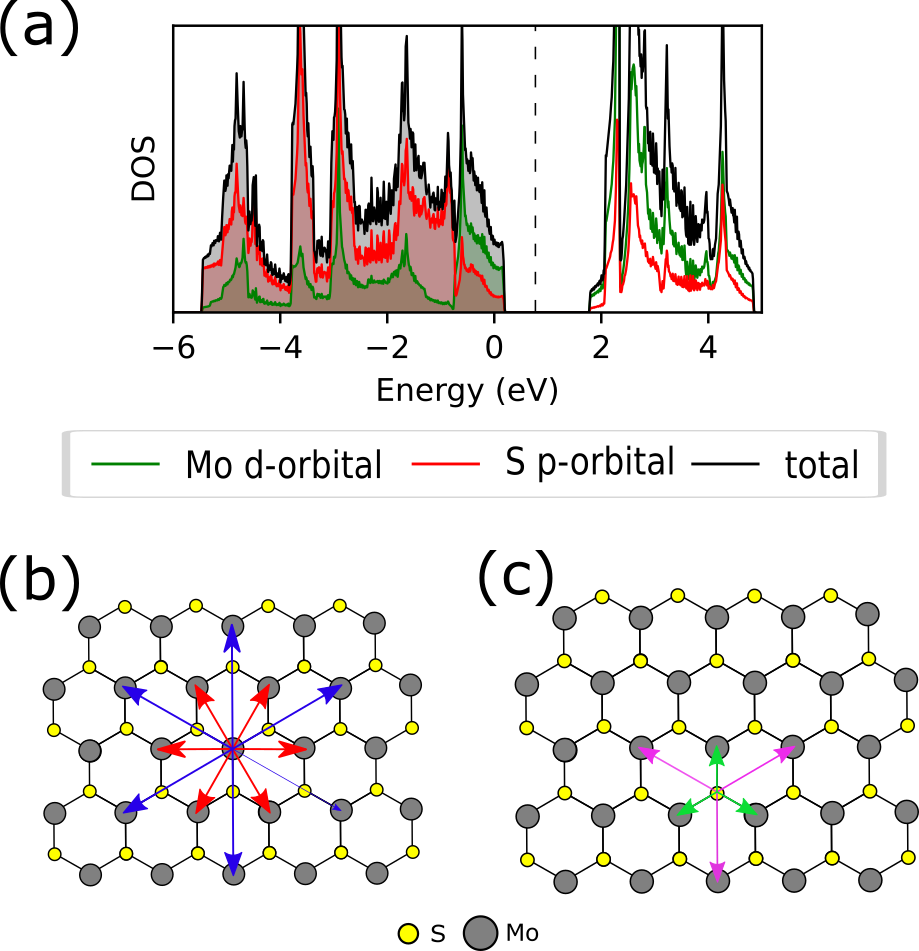}
    \caption{(a) Projected density of states (PDOS) of \ch{MoS2} from DFT calculation. States near the band edge are composed primarily of Mo $d$ orbitals, with smaller but still significant contributions from S $p$ orbitals. Schematic of (b) Mo--Mo and (c) Mo--S nearest-neighbor and next-nearest-neighbor hopping terms considered in tight-binding models.}
    \protect\label{fig:pdos}
\end{figure}

To construct the Wannier function based model, maximally localized Wannier functions were generated using the Wannier90 code~\cite{Marzari97p12847, Marzari12p1419, Mostofi14p2309, Pizzi20p165902}. The band-edge states of \ch{MoS2} are composed primarily of Mo $4d$ orbitals and to a lesser extent S $3p$ orbitals (Fig.~\ref{fig:pdos}a). Therefore, for the five-band model, the Mo $4d$ orbitals were used as initial projectors, and both the Mo $4d$ and S $3p$ orbitals were used as initial projectors for the eleven-band model.
The disentanglement and frozen energy windows were chosen to be \qtyrange{-0.75}{8}{\eV} and \qtyrange{-0.74}{8}{\eV} respectively for the five-band model and \qtyrange{-6}{8}{\eV} and \qtyrange{-6}{5}{\eV} for the eleven-band model.
Tight-binding Hamiltonians were constructed from the maximally localized Wannier functions using the PythTB package and retaining inter-site hopping terms larger than \qty{10}{\meV}~\cite{PythTB}. We define our Bloch states $\left|n\bm{k}\right\rangle$ and tight-binding basis functions $\left|\alpha\bm{R}\right\rangle$ as
\begin{equation}
\left| n\bm{k}\right\rangle
= \sum_{\alpha} C_{n\alpha}(\bm{k}) \left|\alpha\bm{k}\right\rangle
= \sum_{\alpha\mathbf{R}}
e^{i\mathbf{k}\cdot \mathbf{R}} C_{n\alpha}(\bm{k})
\left|\alpha\mathbf{R}\right\rangle
\end{equation}
where $\bm{R}$ indexes the unit cell and $C_{n\alpha}$ are the coefficients for a Bloch wavefunction in the tight-binding basis.

Band structures, dielectric functions, and shift current susceptibility were calculated for both Slater-Koster and Wannier tight-binding models. The shift current was calculated according to Eq.~\ref{eq:shift} using the adaptation to tight-binding models outlined in Ref.~\onlinecite{Dai23pl201201}.
Tight-binding calculations were performed on a $\Gamma$-centered \numproduct{400 x 400} $k$-point mesh with a broadening of \qty{20}{\milli\eV}.
For analysis of quantities over the full Brillouin zone, velocity matrix elements were computed on a $40 \times 40$ k-point mesh. The velocity matrix element between bands $n$ and $m$ is defined as:
\begin{multline}
\label{eq:vmat}
v_{nm} = \frac{p_{nm}}{m_0} = \frac{i}{\hbar}[\hat{H},\mathbf{r}] \\
= \frac{1}{\hbar} \sum_{\alpha, \beta}
C^*_{n\beta}(\bm{k}) C_{m\alpha}(\bm{k})\,
\nabla_{\mathbf{k}}\langle\beta\mathbf{k}|H| \alpha\mathbf{k}\rangle  \\
+ \frac{i}{\hbar} \left\{E_{n\mathbf{k}}-E_{m\mathbf{k}}\right\}
\sum_{\alpha, \beta}
C^*_{n\beta}(\bm{k}) C_{m\alpha}(\bm{k}) \bm{d}_{\beta\alpha} 
\end{multline}
where $\bm{d}_{\beta\alpha}=\left\langle\beta\bm{0}|\bm{r}|\alpha\bm{0}\right\rangle$ is the ``intra atomic'' interaction term described in Ref.~\cite{Pedersen01p201101}. Because we are working in the tight-binding and localized Wannier function approximations, we assume the position operator is diagonal in the tight-binding basis and do not consider these terms.
These treatments are only approximate for systems that include nonlocal terms, including the pseudopotential in \emph{ab initio} calculations. However, nonlocal terms are not present in tight binding models, and the discrepancy arising from the pseudopotential is typically small~\cite{Ibanez18p245143}.

\section{Results and Discussion}

\subsection{DFT Optoelectronic Responses}
\label{sec:DFT}

\begin{figure}
    \centering \includegraphics{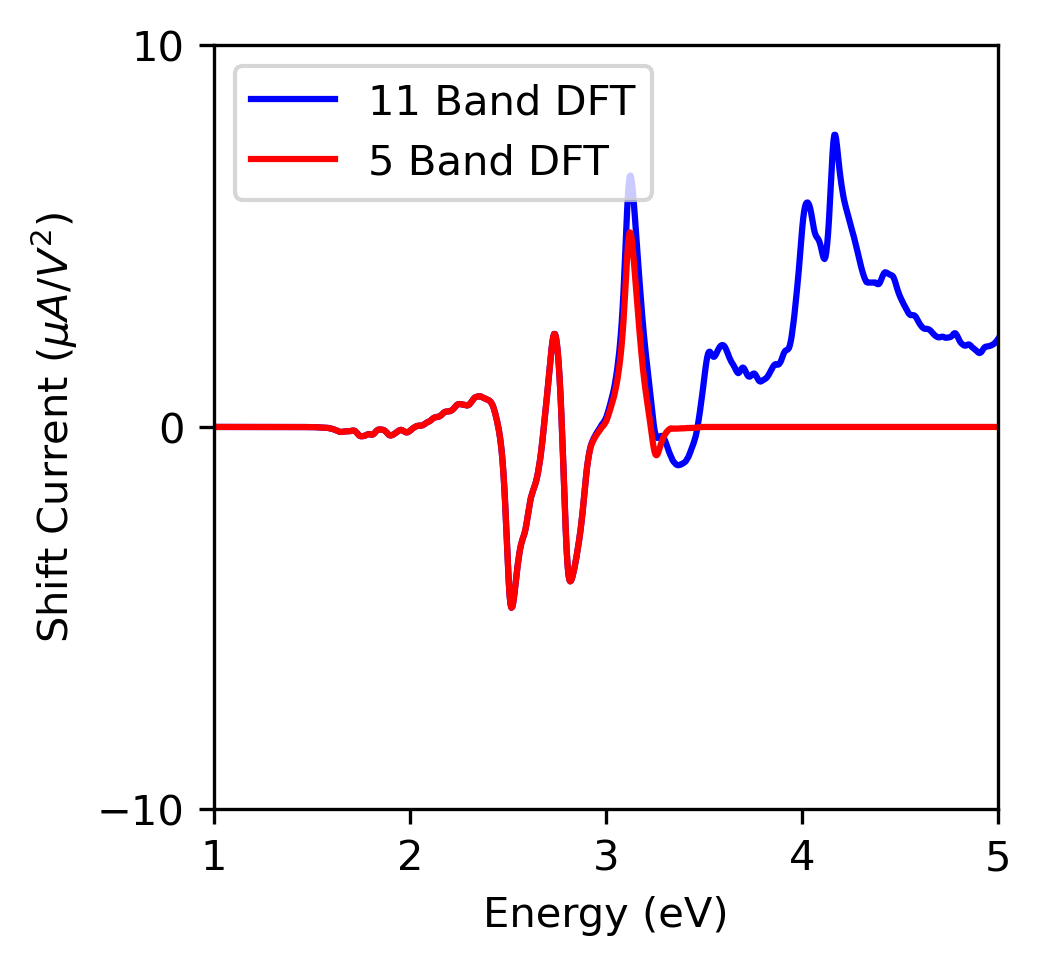}
    \caption{Comparison of the $yyY$ component of the DFT shift current responses for calculations restricted to 5 band and 11 bands.}
    \protect\label{fig:dftbands}
\end{figure}

We first consider the shift current and dielectric tensors as computed from \emph{ab initio} DFT and compare a complete description including contributions from all bands to calculations where the sum over bands is restricted to consider only certain transitions. The results from the restricted calculations can be directly compared with tight-binding models that consider the same subset of bands.
We consider a five-band restriction that includes the highest valence band and the four lowest conduction bands and an eleven-band restriction that includes the seven highest valence bands and the four lowest conduction bands (from about \qtyrange{-6}{5}{\eV}).
Both the five- and eleven-band restrictions describe isolated manifolds and so can be Wannierized easily. The five-band model is compared with existing tight-binding models of TMDs based on \ch{Mo} $d$ orbitals~\cite{Wu15p075310,Liu13p085433}.
The form of the calculated optical properties including the imaginary dielectric function and the shift current (Eq.~\ref{eq:shift}) is resonant, and the DFT wavefunctions are unaffected by the absence of higher-lying bands, so band restrictions do not change the shape of these spectra for light frequencies that remain within the included bands (Fig.~\ref{fig:dftbands}).
This is not the case for properties that require an explicit sum over the conduction band manifold (including many formulations of second-harmonic generation~\cite{Sharma04p128}).
In other models that will be discussed, where models are restricted by limiting the number of real-space orbitals, band truncation can have more significant effects.
These results allow us to compare tight-binding models to DFT results containing an equivalent number of bands, isolating the inaccuracies contributed uniquely by the tight-binding description, and not by the band truncation.

\subsection{Slater-Koster Optoelectronic Responses}
\label{sec:SK}

\begin{figure}
\centering
\includegraphics{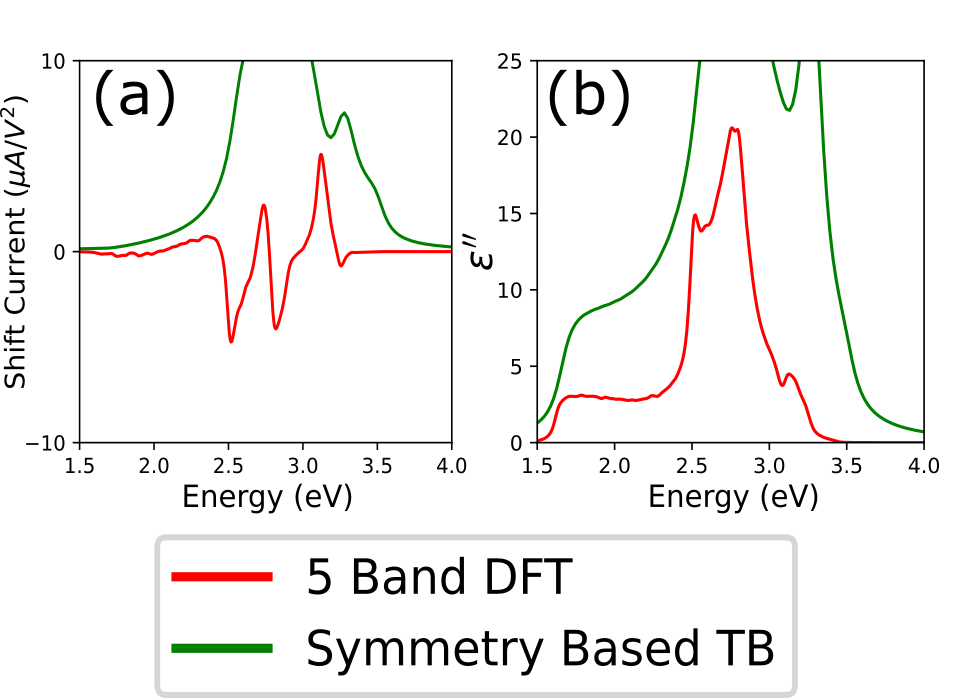}
\caption{(a) Shift current and (b) imaginary dielectric tensor calculated from five bands using DFT and the Slater-Koster model of Ref.~\onlinecite{Wu15p075310}.}
\label{fig:SK}
\end{figure}

In the Slater-Koster method of parameterizing tight-binding models, the form of the Hamiltonian matrix elements is derived analytically from the symmetry and geometry of the system according to a few independent parameters.
These parameters are then determined numerically by a least squares fitting of the algebraic expressions to the results of a first-principles calculation such as DFT~\cite{Slater54p1498}.

Considering a five-band model system composed of all Mo 4$d$ orbitals initially appears to be a reasonable approximation, as the orbital-projected density of states (Fig.~\ref{fig:pdos}a) shows that both the valence band maximum and low-lying conduction bands have primarily Mo 4$d$ character. Many previous works have followed this observation to motivate Slater-Koster tight-binding models using only the Mo sites~\cite{Liu13p085433,Wu15p075310}.
Like other parameterizations, the five-band third-nearest-neighbor model of Ref.~\onlinecite{Wu15p075310} reproduces the band structure and density of states with high accuracy, with an RMSE compared to the DFT band structure of \qty{95}{\milli\eV}, indicating fairly close agreement~\cite{Prandini18p72}.
However, this accuracy does not extend to the optical properties---the dielectric function and shift current calculated from this tight-binding model (Fig.~\ref{fig:SK}) do not agree with the properties predicted using the DFT wavefunctions directly. As both the dielectric tensor and the shift current are gauge invariant both globally and at each $k$-point (Supplementary Information), this discrepancy indicates an inaccuracy that influences physical observables.
While the energy of the onset is similar (which is expected, as the Slater-Koster model exactly reproduces the band gap), the shift current spectrum from the tight-binding model lacks several key features seen in DFT, and the spectra have considerably different magnitudes.
Therefore, good band structure agreement is not sufficient to show that a tight-binding model will yield accurate optical properties.

\subsection{Impact of Band Restriction on Wannier Function Optoelectronic Responses}
\label{sec:bandrestrict}

\begin{table}
\centering
\caption{
Accuracy of electronic and optical properties calculated using different tight binding models and different hopping ranges. Band structure error is expressed as an RMSE in \unit{\meV}. Spread refers to the average spread of the Wannier functions used in each model in \unit{\angstrom\squared}.}
\label{tab:accuracy}
\begin{ruledtabular}
\begin{tabular}{lccc}
&& NNN & Beyond NNN \\
\hline
\multirow{4}{*}{Slater-Koster} & Spread & \multirow{4}{*}{-} & N/A  \\
& Band RMSE & & \qty{95}{\meV} \\
& $\varepsilon_{xx}$ & & Poor \\
& $\sigma_{yyY}$ & & Poor \\
\hline
\multirow{4}{*}{Wannier 5-band} & Spread & \qty{5.002}{\angstrom\squared} & \qty{5.002}{\angstrom\squared} \\
& Band RMSE & \qty{141}{\meV} & \qty{38}{\meV} \\
& $\varepsilon_{xx}$ & Qualitative & Qualitative \\
& $\sigma_{yyY}$ & Poor & Qualitative \\
\hline
\multirow{4}{*}{Wannier 11-band} & Spread & \qty{1.676}{\angstrom\squared} & \qty{1.676}{\angstrom\squared}  \\
& Band RMSE & \qty{104}{\meV} & \qty{40}{\meV} \\
& $\varepsilon_{xx}$ & Quantitative & Quantitative \\
& $\sigma_{yyY}$ & Qualitative & Quantitative \\
\end{tabular}
\end{ruledtabular}
\end{table}

\begin{figure}
\centering
\includegraphics{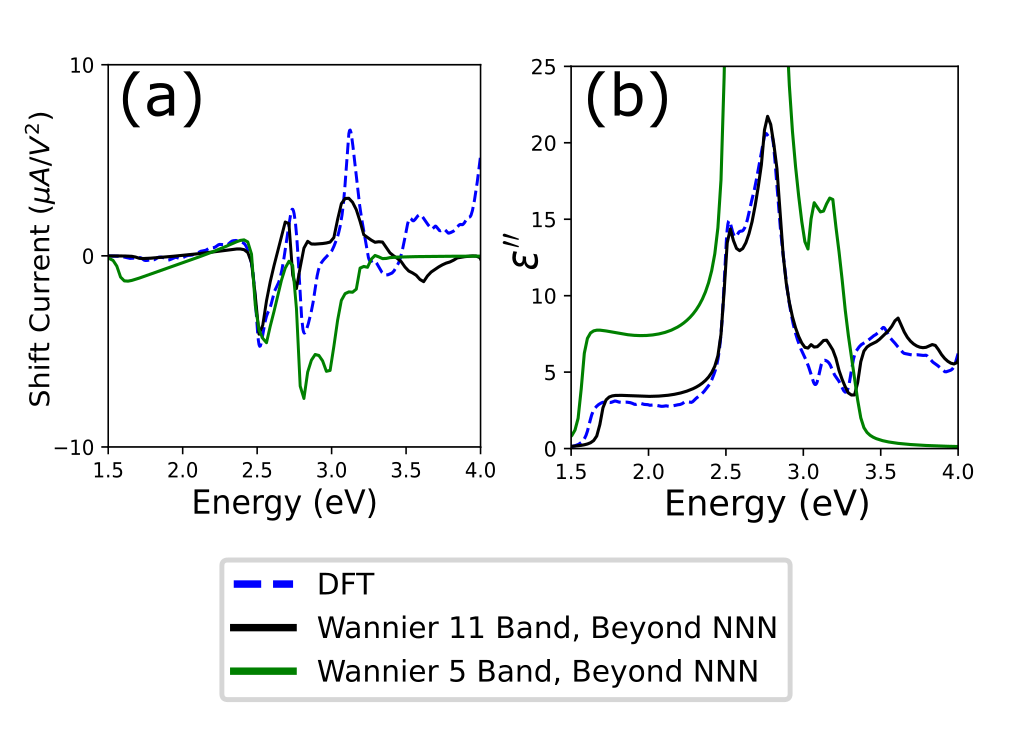}
\caption{(a) Shift current, and (b) imaginary dielectric tensor restricted to eleven bands calculated from DFT and Wannier tight binding models with five and eleven bands and beyond next nearest-neighbor hoppings.}
\label{fig:wann_tb}
\end{figure}

An alternative representation of the system can be achieved by constructing a set of Wannier functions that are well-localized in real space~\cite{Marzari97p12847, Marzari12p1419}. The Hamiltonian matrix elements projected into the basis of these real-space orbitals can then be used as hopping terms in a tight-binding model.
A five-band tight-binding model constructed in this way has better agreement with DFT band structure than the Slater-Koster model, with an RMSE of \qty{38}{\milli\eV} (Table~\ref{tab:accuracy}). The qualitative features of the shift current and dielectric responses are also better reproduced, but the magnitudes are still not comparable to DFT (Fig.~\ref{fig:wann_tb}).

The low quality of the five-band Wannier model arises because a real space model including only Mo 4$d$-like orbitals is not a suitably accurate description of the true wavefunction character.
To mitigate this shortcoming, we consider an eleven-band tight-binding model that also includes the sulfur $3p$ orbitals, which make appreciable contributions at the band edges (Fig.~\ref{fig:pdos}).
This model fits the DFT band structure well, with an RMSE of \qty{40}{\milli\eV}.
Furthermore, the eleven-band model has good agreement with the DFT shift current spectrum in the visible range, and a good agreement with the dielectric tensor in an even larger range (Fig.~\ref{fig:wann_tb}).
One potential reason that the shift current does not agree with DFT quite as well as the dielectric function does is that any non-diagonal contributions to the position operator are ignored in our tight-binding formalism. This is only approximately true for many tight-binding models, so the fidelity of the optical property calculations can be further improved by taking these additional overlaps between orbitals into account~\cite{Ibanez22p070}.
From these results we see that band structure agreement is necessary, but not sufficient, for accurate optical calculations. All models discussed show reasonable band structure agreement, but the Wannierization procedure, unlike the Slater-Koster formalism, seeks to reproduce the Bloch states in addition to the band energies. This improvement carries over to the optical property calculations, which are explicitly wavefunction dependent.

\subsection{Impact of Hopping Range on Wannier Function Optoelectronic Responses}
\label{sec:hoppings}

\begin{figure}
\centering
\includegraphics{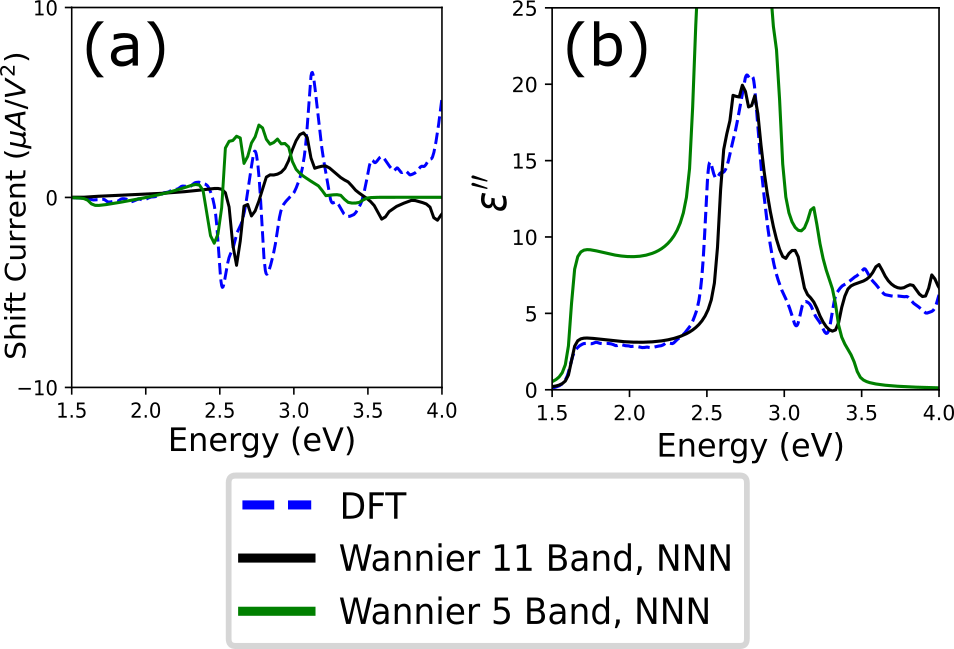}
\caption{(a) Shift current and (b) dielectric tensor calculated from DFT restricted to eleven bands compared to the same quantities calculated with Wannier Function models truncated to next nearest-neighbor (NNN) hoppings.}
\label{fig:Hopping_nnn}
\end{figure}

Wannier interpolation formally considers interactions between Wannier functions with arbitrarily large separations, but practically, the magnitude of these interactions decays quickly with distance due to the localization of the Wannier functions. The Wannier functions were sufficiently localized in each of the Wannier Function models considered here, with a spread of \qty{5.002}{\angstrom^2} for the five-band model and \qty{1.646}{\angstrom^2} for the eleven-band model.
This is a key component of the efficiency and interpretability of Wannier tight-binding models.
Tight binding models often neglect hoppings beyond next-nearest neighbor (NNN), though the model considered here from Ref.~\onlinecite{Wu15p075310} includes third-nearest neighbor hoppings.
We next investigate the impact of neglecting long range hoppings on the optoelectronic response by constructing Wannier tight-binding models that only include interactions through NNN hopping.
Unsurprisingly, this negatively impacts the band structure agreement, with an RMSE of \qty{141}{\milli\eV} for the five-band NNN model and \qty{104}{\milli\eV} for the eleven-band NNN model, while discarding more significant terms to make nearest-neighbor models incurs correspondingly larger errors.
Although the band-structure agreement is worse than for the Wannier model with unrestricted hopping, the NNN model is still comparable to the Slater-Koster model with a similar hopping range.

The optical properties calculated using these restricted models do not agree with either DFT or the unmodified Wannier models.
Despite the comparable or worse band structure agreement compared to the Slater-Koster model, the restricted Wannier models better reproduce the qualitative features and approximate magnitude of the optical responses, including a relatively low magnitude near the band gap energy and several changes in shift current direction with increasing photon frequency.
The eleven-band NNN model shows quite close agreement in magnitude for both the shift current and the dielectric function, showing that well-localized Wannier models can be constructed that provide a reasonable description of the optical properties.

\subsection{Origin of Model Inaccuracy}
\label{sec:BZ}

\begin{figure}
\centering
\includegraphics{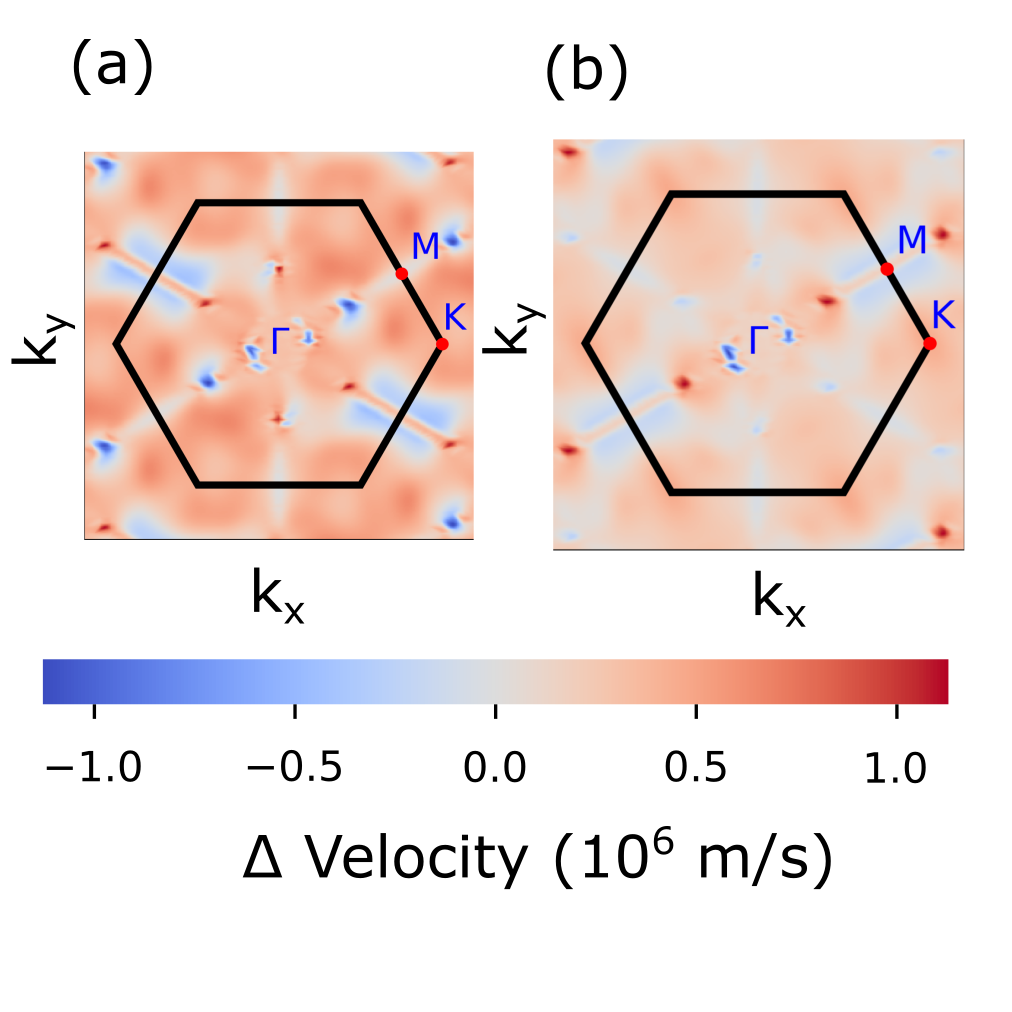}
\caption{Difference between the five-band and eleven-band Wannier beyond NNN models for the the interband velocity matrix element between the highest valence band and lowest conduction bands. The component of velocity along the first reciprocal lattice vector is shown in (a) and the component along the second reciprocal lattice vector is shown in (b). The Brillouin zone is indicated with the dotted line. The positive region around the $K$ high-symmetry point suggests that the five-band model overestimates optical response at the onset.
Also visible are small regions with a very large discrepancy between models. These originate from band crossings that occur at slightly different locations in the Brillouin zone. However, due to their small area, they have less impact than the discrepancies around the $K$ point.
}
\label{fig:vmatgrid}
\end{figure}

To gain better insight into shift current and dielectric properties, which are Brillouin zone averaged quantities, we analyze the discrepancies between the different tight-binding models throughout the entire Brillouin zone.
Like in the band structures discussed above, the band energies of the Slater-Koster model and the two Wannier tight-binding models agree with DFT (see Supplementary Information). While there are slight variations in the quantitative agreement, they do not entirely explain the inconsistencies in optical responses between models, because the overall energy landscapes have very similar behavior.

We next examine the interband velocity matrix elements along both reciprocal space axes using a dense k-point grid across the Brillouin zone. Variations in the calculated interband velocities could indicate differences in optical responses, including the shift current and dielectric tensor. Our analysis focuses on the magnitude of the interband velocity between the top valence band and the bottom conduction band, as this parameter plays a crucial role in determining the optical response at the onset of absorption. The dielectric function depends directly on the interband velocity, while the shift current depends on both the interband velocity and on the derivative of the phase of the velocity matrix, through the shift vector (Eq.~\ref{eq:shift}). Deviations in the interband velocity will therefore have an effect on both the dielectric and shift current responses.
Our comparison shows that the five-band Wannier beyond NNN model overestimates the interband velocity around the $K$ high symmetry point when compared to the eleven-band model (Fig.~\ref{fig:vmatgrid}). This overestimation of the velocity matrix at the band gap can explain the overestimation of both the shift current and dielectric function at low photon energy.

The difference between interband velocities but not band energies is a clear indication that the quality of the calculated optoelectronic response is not guaranteed by a faithful reproduction of the band structure.
One key quantity included in the calculation for interband velocity but not for band energies is the eigenvectors of the tight-binding Hamiltonian. These matrix elements are not only used for interband velocity but for other quantities such as the Berry connection, which is vital for BPVE shift current. 
The discrepancies in velocity matrix elements between five and eleven-band models despite very similar band energies suggests that the Bloch states are better captured in the eleven-band Wannier beyond NNN model.
This highlights the importance of both good band structure and wavefunction agreement in a tight-binding model used to calculate optical responses. The results presented here suggest that a careful examination of quantities beyond band energies is needed when gauging the accuracy of tight-binding models.

\section{Conclusion}

From these results, we deduce several general features that facilitate accurate calculation of optical properties using tight-binding models.
First, we establish that good band structure agreement is necessary for accurate optical properties, as seen from the worse description of shift current and dielectric response in models where the hopping terms were truncated spatially, despite the fact that similar models without truncation performed well.
However, while accurate description of the band structure is necessary, it is not sufficient for accurate modeling of the optical response---the Slater-Koster and five-band Wannier models tested performed poorly, despite faithfully reproducing band energies, because they did not accurately represent the interband velocity matrix elements.
This has important implications for designing tight-binding models, as band agreement is a commonly used diagnostic to validate model quality before calculating other properties. Instead, models should be tested on the property of interest. As these validations can be performed on small unit cells before scaling the model to larger systems or systems that include additional interactions, they do not mitigate the advantages of using a tight-binding model.

In the case of optical properties, we show that although not all models that fit the band structure yield good results for optical properties, Wannier tight-binding models can. This is because Wannier models explicitly fit both the band energies and the Bloch functions, yielding a better description of the wavefunction character, which is further improved when enough basis functions to represent all chemically relevant orbitals are included in the model.
This allows the model to accurately represent the delocalization and covalancy of the system. These properties impact shift current~\cite{Tan19p084002}, and matrix element agreement explains why a Slater-Koster model does not perform as well as a Wannier tight-binding model restricted to include a similar number of bands and hopping range.
However, it also indicates that many common approximations used to make model systems analytically simple, such as considering only one type of orbital or only very short-range interactions, can have a deleterious effect on the quantitative agreement with first principles.
As tight-binding models are often used to model very large systems or complex responses that can not be directly benchmarked against \emph{ab initio} DFT calculations, the insight outlined here will help guide model design and ensure accurate results.

\section{Acknowledgements}

We acknowledge valuable discussions with Dr. Zhenbang Dai, Dr. Zhenyao Fang, Prof. Youngkuk Kim, and Prof. Nicola Marzari.
This work was supported by the U.S. Department of Energy, Office of Science, Basic Energy Sciences, under Award No. DE-SC0024942.
A. G. acknowledges financial support from the Vagelos Integrated Program in Energy Research at the University of Pennsylvania. 
Computational support was provided by the National Energy Research Scientific Computing Center (NERSC), a U.S. Department of Energy, Office of Science User Facility located at Lawrence Berkeley National Laboratory, operated under Contract No. DE-AC02-05CH11231.


\end{document}


\title{Supporting Information for: \\[0.5em] Choosing Tight-Binding Models for Accurate Optoelectronic Responses}

\author{
Andreas Ghosh
}
\affiliation{
Department of Chemistry, University of Pennsylvania, Philadelphia, PA 19104, USA
}

\author{
Aaron M. Schankler
}
\affiliation{
Department of Chemistry, University of Pennsylvania, Philadelphia, PA 19104, USA
}

\author{
Andrew M. Rappe
}
\affiliation{
Department of Chemistry, University of Pennsylvania, Philadelphia, PA 19104, USA
}
\email{rappe@sas.upenn.edu}

\maketitle
\tableofcontents
\counterwithin{figure}{section}
\counterwithin{equation}{section}
\counterwithin{table}{section}

\clearpage

\section{Band Structures of Tight-Binding Models}

Here we present the band structures for each tight-binding model referenced in the main text.

\subsection{Slater-Koster Band Structure}

First we present the Slater-Koster band structure for monolayer \ch{MoS2} presented by Ref.(insert ref number). This third nearest neighbor tight-binding model reproduces the DFT bands faithfully (with an RMSE of \qty{95}{\meV}), with some minor deviations far from the band edges.

\begin{figure}[h]
\centering
\includegraphics{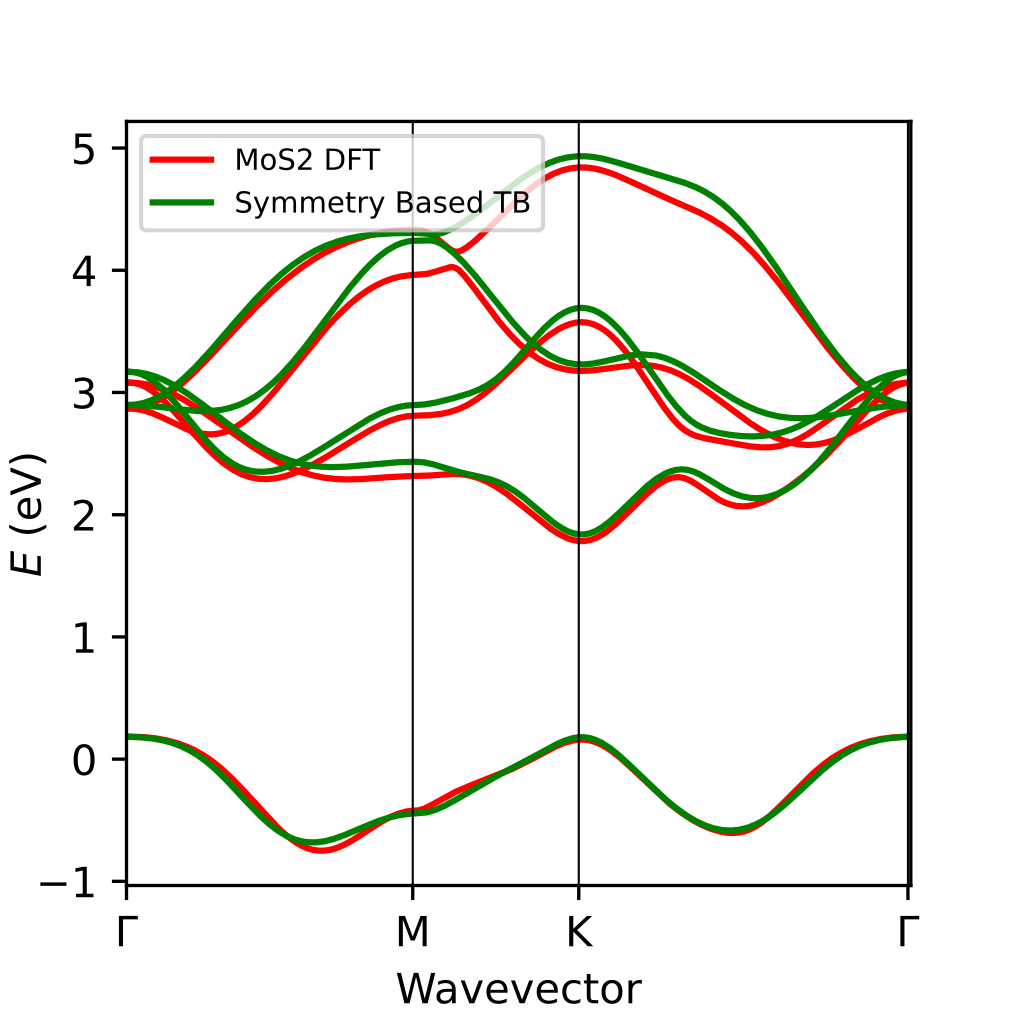}
\caption{DFT band structure of \ch{MoS2} compared to the 5 band symmetry-based tight-binding model. This third-nearest neighbor model has an RMSE of \qty{95}{\meV}.}
\label{fig:dftsymm}
\end{figure}

\clearpage
\subsection{Impact of Band Restrictions on Band Structure}

We now examine the band structures of the 5 band and 11 band Wannier function tight-binding models without NNN truncation. It was established that the 11 band Wannier function model had significantly better optoelectronic responses, despite a slightly larger band structure RMSE of \qty{40}{\meV} (compare to an RMSE of \qty{38}{\meV} in the 5 band case). The band structures of these models are almost indistinguishable for the 5 highest energy bands that are common to both models. This further establishes that neither a visual nor numerical inspection of the band structure is sufficient to assess the quality of a tight-binding model for optical calculations.

\begin{figure}[h]
\centering
\includegraphics{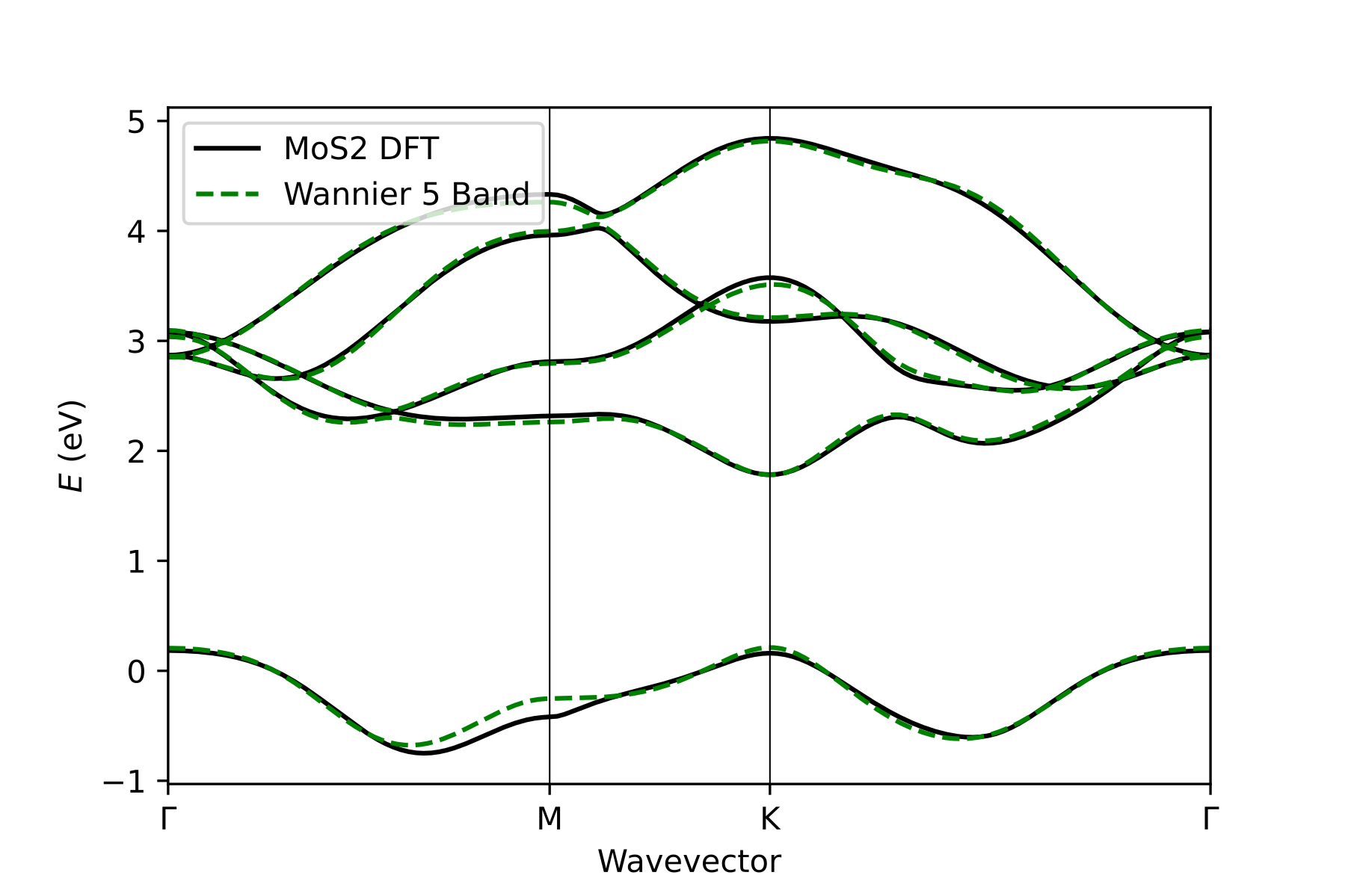}
\caption{DFT band structure of \ch{MoS2} compared to Wannier five-band beyond NNN tight-binding model. This band structure had an RMSE of \qty{38}{\meV}.}
\label{fig:5bandwannier}
\end{figure}

\begin{figure}
\centering
\includegraphics{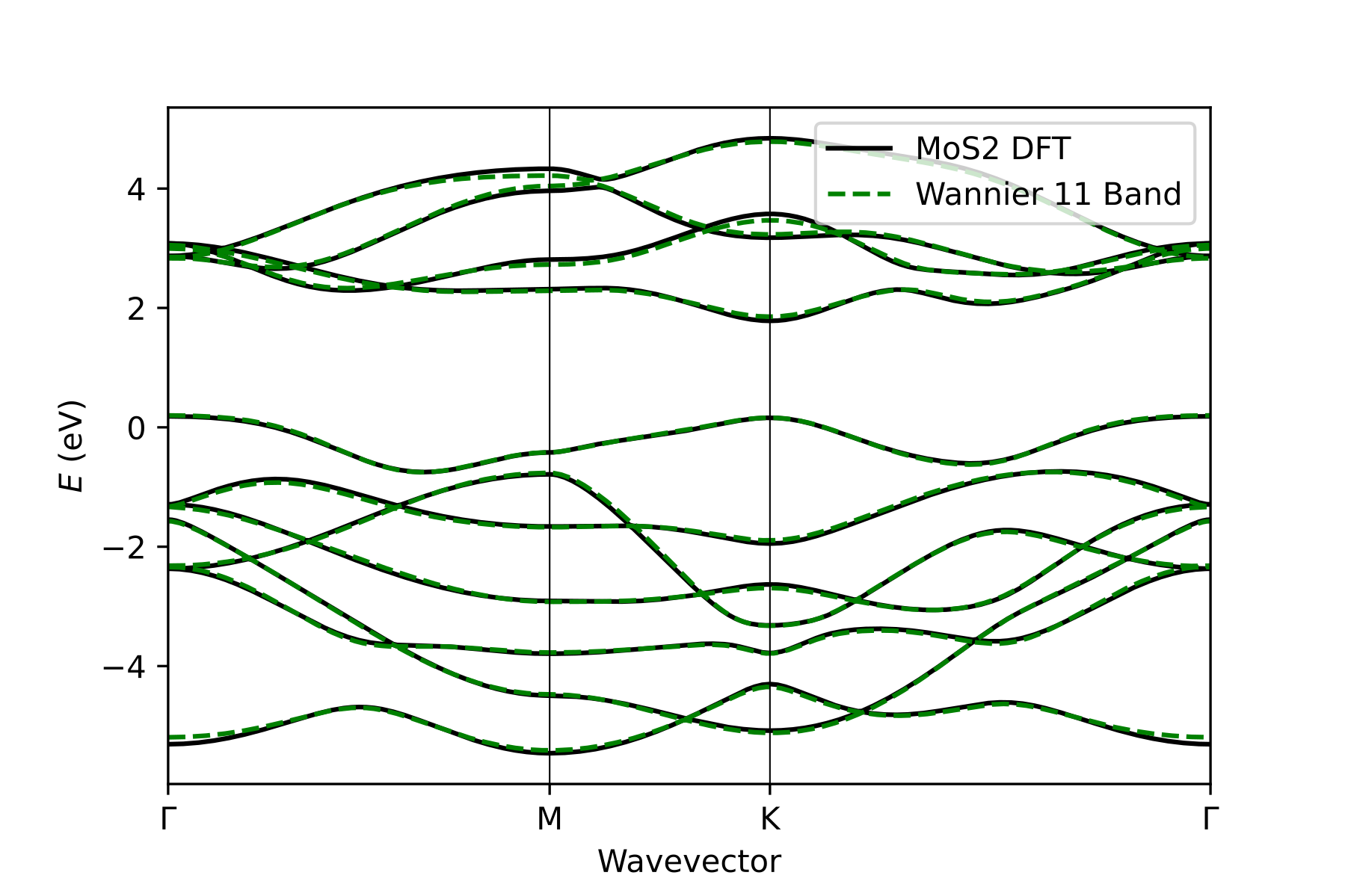}
\caption{DFT band structure of \ch{MoS2} compared to Wannier eleven-band beyond NNN tight-binding model. This band structure had an RMSE of \qty{40}{\meV} for the 5 highest energy bands.}
\label{fig:11bandwannier}
\end{figure}

\clearpage
\subsection{Impact of Hopping Range on Band Structure}

In the main text we find, like many other papers on tight-binding models, that significantly truncating the spatial extent of a tight-binding model significantly worsens its ability to predict optical properties. This can be seen in the significantly larger band structure RMSE's of both models (\qty{104}{\meV} in the 11 band case and \qty{141}{\meV} in the 5 band case). Here, the quality (or lack thereof) of the band structure relative to DFT is indicative of optoelectronic response.

\begin{figure}[h]
\centering
\includegraphics{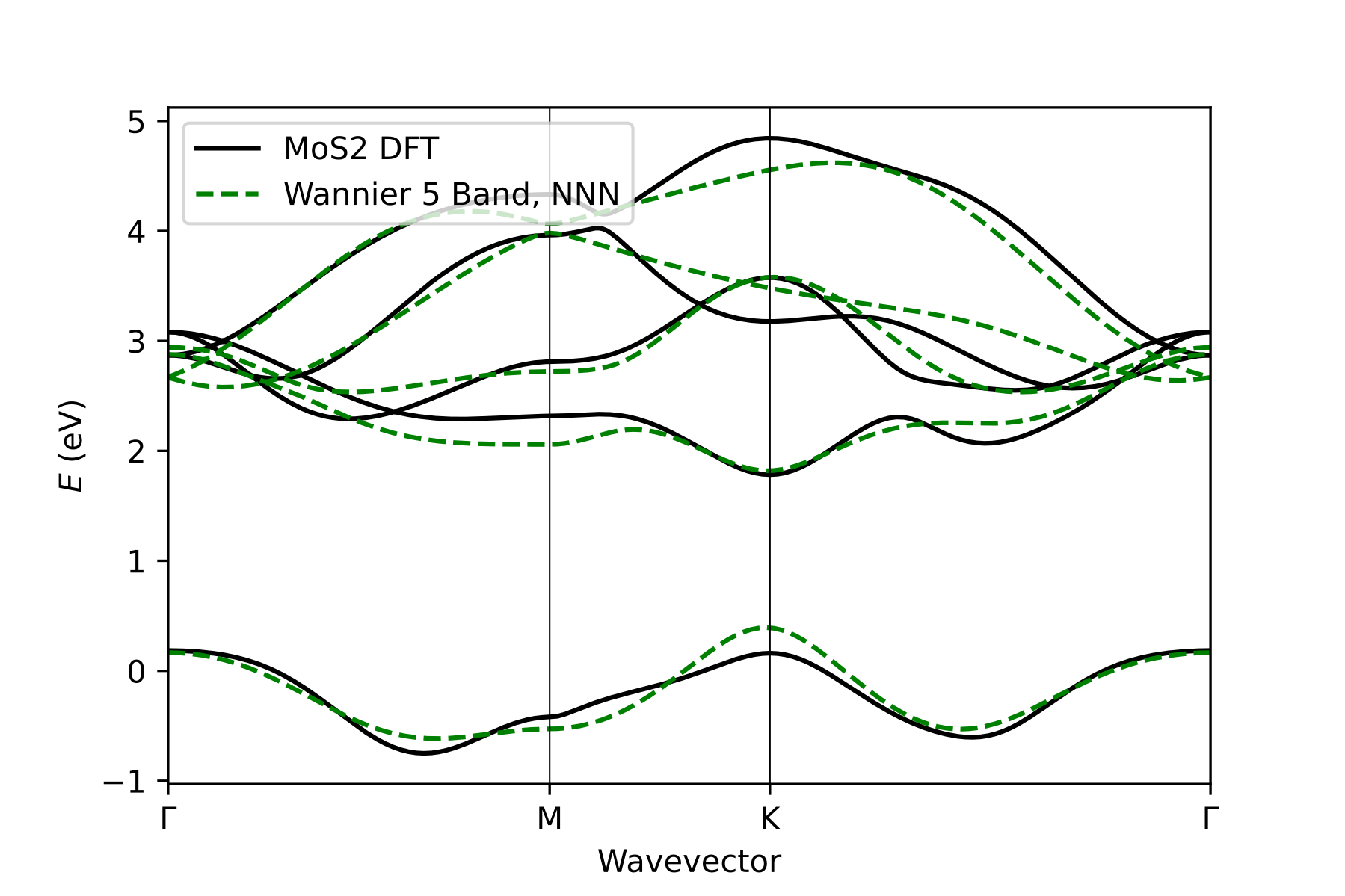}
\caption{DFT band structure of \ch{MoS2} compared to 5 band Wannier tight-binding model with hoppings truncated to NNN. This band structure had an RMSE of \qty{141}{\meV}.}
\label{fig:5bandwannierNNN}
\end{figure}

\begin{figure}
\centering
\includegraphics{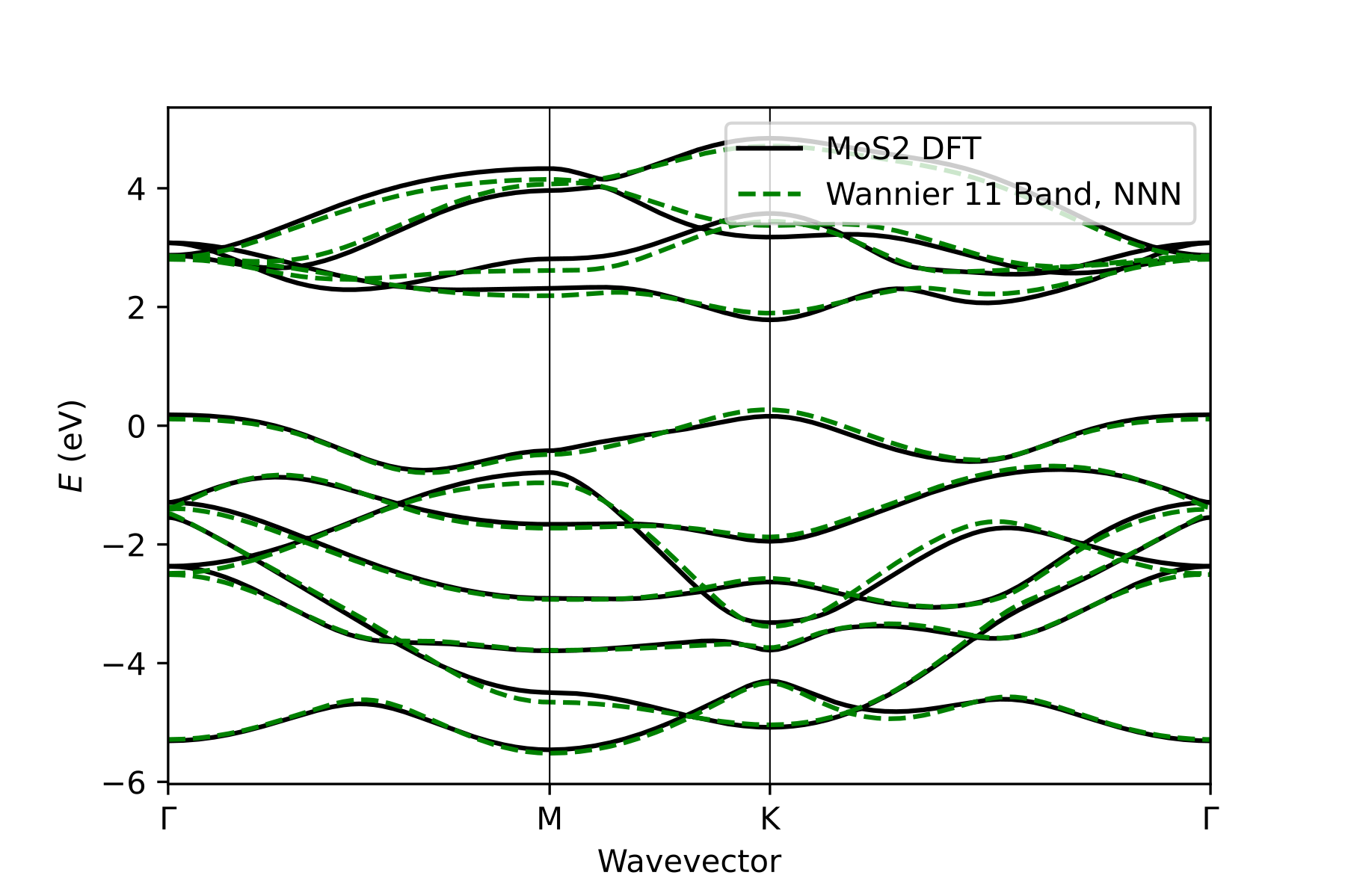}
\caption{DFT band structure of \ch{MoS2} compared to 11 band Wannier tight-binding model with hoppings truncated to NNN. This band structure had an RMSE of \qty{104}{\meV} for the 5 highest energy bands.}
\label{fig:11bandwannierNNN}
\end{figure}

\clearpage
\section{Brillouin Zone Energy Analysis}

Here, we include results from a $20 \times 20$ Monkhorst-Pack grid of energies calculated for the five bands of interest from each model. We can see that the Slater-Koster model performs the worst, and the five and eleven-band models have comparable error versus DFT.

\begin{table}[h]
\centering
\caption{
Band structure error between DFT and tight-binding models. Band structure error is expressed as an RMSE in \unit{\meV}. Band 1 is the maximum valence band and band 5 is the maximum conduction band.}
\protect\label{tab:bands}

\begin{tabular}{SSSS}
 \toprule
& \multicolumn{3}{c}{Error from DFT Bands} \\
 \cmidrule(lr){2-4}
{Band Index} & {5 Band} & {11 Band} & {Slater-Koster}\\
 \midrule
 1 & 42 & 14 & 29 \\
 2 & 28 & 28 & 58 \\
 3 & 31 & 33 & 116 \\
 4 & 27 & 41 & 116 \\
 5 & 26 & 40 & 129 \\
 \hline
 \textbf{Avg.}& 30.8  & 31.2   & 89.6\\
\bottomrule
\end{tabular}
\end{table}

\clearpage
\section{Comparison to Shift Current Implementation in Wannier90}

A direct calculation of the shift current can be performed in Wannier90 without the need for PythTB or similar codes for tight-binding models (see below). This calculation produces different results from ours because of the direct incorporation of nonlocal potential terms in the velocity matrix and off-diagonal position matrix elements in the Wannier90 shift current code. However, the lack of these elements in our implementation does not detract from our claim that wavefunction agreement must be verified in tight-binding models. The discrepancy between shift current responses between tight-binding models discussed in this letter persist without these additional terms.
The shift current calculation below used a $400 \times 400 \times 1$ k-point grid with a broadening of \qty{80}{meV}. This broadening was chosen for the spectra to match qualitatively.

\begin{figure}[h]
\centering
\includegraphics{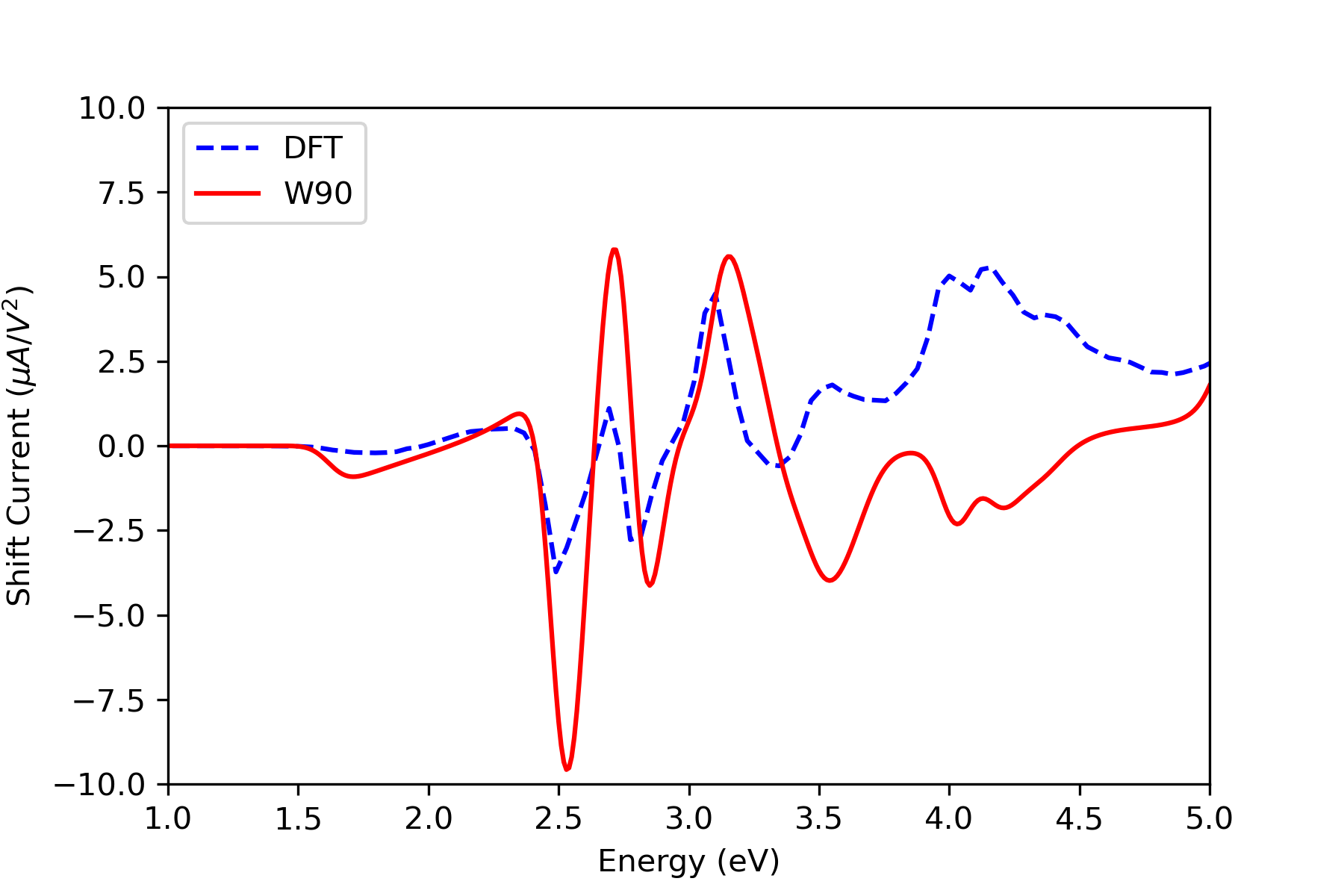}
\caption{DFT shift current compared to shift current calculated directly with Wannier90. This calculation is qualitatively similar to the one presented in this letter, but is slightly different due to calculation methodology.}
\label{fig:dftw90}
\end{figure}

\clearpage
\section{Gauge Invariance of Shift Current Calculations}

At first inspection, the significant differences in shift current spectra between different models suggest a numerical gauge covariance. These concerns are valid as the Berry Connection for the shift vector is a gauge covariant quantity. However, shift current is overall gauge invariant. Here, we describe the numerical implementation that ensures gauge invariant shift current calculations. To start, the shift vector $R$ can be written as follows:
\begin{equation}
    R=-i \frac{\partial \ln \left\langle n^{\prime \prime} k|\hat{v}| n^{\prime} k\right\rangle}{\partial k}-\left[A^{\prime \prime}(k)-A^{\prime}(k)\right]
\end{equation}
We use the following definition of the Berry Connection:
\begin{equation}
    A(k)=i\left\langle u_{k} | \frac{\partial u_{k}}{\partial k}\right\rangle=i \lim _{\Delta k \rightarrow 0} \frac{1}{\Delta k} \ln \left\langle u(k) \mid u(k+\Delta k)\right\rangle
\end{equation}
We abbreviate the ket $|u_{nk}\rangle$ as $|nk\rangle$ and use the following discretization:
\begin{equation}
\begin{aligned}
R= & -i \frac{1}{\Delta k}\left(\ln \left\langle n^{\prime \prime} k+\Delta k|\hat{v}| n^{\prime} k+\Delta k\right\rangle-\ln \left\langle n^{\prime \prime} k|\hat{v}| n^{\prime} k\right\rangle\right) \\
& -\left[i \frac{1}{\Delta k}\left\langle n^{\prime \prime} k \mid n^{\prime \prime} k\right\rangle\left(\ln \left\langle n^{\prime \prime} k \mid n^{\prime \prime} k+\Delta k\right\rangle-\ln \left\langle n^{\prime \prime} k \mid n^{\prime \prime} k\right\rangle\right)\right. \\
& \left.-i \frac{1}{\Delta k}\left\langle n^{\prime} k \mid n^{\prime} k\right\rangle\left(\ln \left\langle n^{\prime} k \mid n^{\prime} k+\Delta k\right\rangle-\ln \left\langle n^{\prime} k \mid n^{\prime} k\right\rangle\right)\right] \\
= & -i \frac{1}{\Delta k} \ln \frac{\left\langle n^{\prime \prime} k+\Delta k|\hat{v}| n^{\prime} k+\Delta k\right\rangle}{\left\langle n^{\prime \prime} k|\hat{v}| n^{\prime} k\right\rangle}-i \frac{1}{\Delta k} \ln \frac{\left\langle n^{\prime \prime} k \mid n^{\prime \prime} k+\Delta k\right\rangle}{\left\langle n^{\prime} k \mid n^{\prime} k+\Delta k\right\rangle} \\
= & -i \frac{1}{\Delta k} \ln \frac{\left\langle n^{\prime \prime} k+\Delta k|\hat{v}| n^{\prime} k+\Delta k\right\rangle\left\langle n^{\prime \prime} k \mid n^{\prime \prime} k+\Delta k\right\rangle}{\left\langle n^{\prime \prime} k|\hat{v}| n^{\prime} k\right\rangle\left\langle n^{\prime} k \mid n^{\prime} k+\Delta k\right\rangle} .
\end{aligned}
\end{equation}
This numerical implementation results in the same $R$ regardless of gauge. To show this, we perform the following change of variables:
\begin{equation*}
\begin{aligned}
    &\left|n^{\prime} k\right\rangle\rightarrow e^{i\phi'_{k}}\left|n^{\prime} k\right\rangle \\
    &\left|n^{\prime\prime} k\right\rangle\rightarrow e^{i\phi''_{k}}\left|n^{\prime\prime} k\right\rangle \\
    &\left|n^{\prime} k+\Delta k\right\rangle\rightarrow e^{i\phi'_{k+\Delta k}}\left|n^{\prime} k+\Delta k\right\rangle \\
    &\left|n^{\prime\prime} k+\Delta k\right\rangle\rightarrow e^{i\phi''_{k+\Delta k}}\left|n^{\prime\prime} k+\Delta k\right\rangle \\
\end{aligned}
\end{equation*}
Now we plug these phases into our discretization:
\begin{equation*}
R= -i \frac{1}{\Delta k} \ln \frac{e^{-i\phi''_{k+\Delta k}}e^{i\phi'_{k+\Delta k}}\left\langle n^{\prime \prime} k+\Delta k|\hat{v}| n^{\prime} k+\Delta k\right\rangle e^{i\phi''_{k+\Delta k}}e^{-i\phi''_{k}}\left\langle n^{\prime \prime} k \mid n^{\prime \prime} k+\Delta k\right\rangle}{e^{-i\phi''_{k}}e^{i\phi'_{k}}\left\langle n^{\prime \prime} k|\hat{v}| n^{\prime} k\right\rangle e^{-i\phi'_{k}}e^{i\phi'_{k+\Delta k}} \left\langle n^{\prime} k \mid n^{\prime} k+\Delta k\right\rangle} .
\end{equation*}
By inspection all these phases cancel.